\documentclass[oldversion]{aa}
\usepackage{graphicx}
\usepackage[authoryear]{natbib}
\usepackage{txfonts}
\begin{document}
   \title{Updating the orbital ephemeris of Her~X-1; rate of decay and 
   eccentricity of the orbit}


   \author{R.~Staubert\inst{1}\and
               D.~Klochkov\inst{1}\and
              J.~Wilms\inst{2,3}
              }

\offprints{staubert@astro.uni-tuebingen.de}

   \institute{Institut f\"ur Astronomie und Astrophysik, 
     University of T\"ubingen, Sand 1, 72076 T\"ubingen, Germany
   \and
   Dr. Karl Remeis-Sternwarte, Astronomisches Institut der
   Universit\"at Erlangen-N\"urnberg, Sternwartstr. 7, 96049 Bamberg, Germany
  \and
  Erlangen Center for Astroparticle Physics (ECAP), Erwin-Rommel-Str. 1, 91058 Erlangen, Germany
   }


   \date{Received 21.01.2009; accepted 30.03.2009}

\abstract{     
     We present an update of the orbital ephemeris of the binary X-ray pulsar Her X-1
      and determine an improved value for the rate of orbital decay. In addition,
      we report the first measurement of the orbital eccentricity.    
      We have analyzed pulse timing data of Her X-1 from X-ray observations by 
      \textsl{RXTE} (\textsl{Rossi X-Ray Timing Explorer}) and \textsl{INTEGRAL} 
      over the period 1996 -- 2007. Accurate pulse arrival times were 
      determined from solar system bary-centered photon arrival times by generating 
      pulse profiles averaged over appropriately short integration times. 
      Applying pulse phase connection techniques, it was possible to determine sufficiently 
      accurate local ephemeris data for seven observation periods distributed over 12 years. 
      Combining the new local $T_{\pi/2}$ values with historical values from the literature 
      we update the orbital ephemeris of Her X-1 to $T_{\pi/2}$ = MJD 46359.871940(6) 
      and $P_{\rm orb}$ = 1.700167590(2)\,d and measure a continuous change 
      of the orbital period of  $dP_{\rm orb}/dt = -(4.85 \pm 0.13) \times 10^{-11} {\rm ss^{-1}}$.  
      For the first time, a value for the eccentricity of the orbit of  Her~X-1 is measured
      to be $\epsilon$ = $(4.2 \pm 0.8) \times 10^{-4}$.
}

   \keywords{stars: binaries: general -- stars: neutron -- X-rays: general -- X-rays: binaries -- 
   X-rays: individuals: Her X-1 -- Ephemerides}

   \maketitle
%

\section{Introduction}
Hercules~X-1 and Centaurus~X-3 were the first two X-ray pulsars discovered 
by the \textsl{Uhuru} satellite in 1972 \citep{Tananbaum_etal72,Giacconi_etal73}.
Since then they have remained among the most studied X-ray sources.
Her~X-1 is a close binary system consisting of an accreting magnetized 
neutron star and the stellar companion HZ~Her \citep[first suggested 
by][]{Liller72}, a main sequence star of spectral type A/F \citep{Crampton74}. 
The mass of the optical companion is $\sim$2.2$M_{\odot}$  \citep{Deeter_etal81} 
which places the system in between high and low mass X-ray binaries. 
The orbital period of the system is 1.700\,d.
Due to the high inclination of the orbital plane ($i \sim 85^{\circ}-88^{\circ}$) 
the X-ray source is regularly eclipsed by the optical 
companion for $\sim$5.5\,h once per orbit. The X-ray
luminosity of the source is  $L_{\rm X}\sim 2\times 10^{37} {\rm erg\,s}^{-1}$ 
for a distance of $\sim$7~kpc \citep{Reynolds_etal97}. The spin period
of the neutron star, represented by strong X-ray pulsations, is 1.237\,s.

The very first \textsl{Uhuru} observations of Her X-1 revealed the
presence of a long-term $\sim$35\,d periodicity which manifests itself mainly 
through the alternation of so-called {\em on} (high X-ray flux) and {\em off} 
(low X-ray flux) states. The 35\,d cycle
contains two {\em on} states -- the {\em main-on} ($\sim$7 orbital periods)
and the {\em short-on} ($\sim$5 orbital cycles) -- separated by $\sim$4--5
orbital cycles. This periodicity in Her~X-1/HZ~Her is usually attributed to a 
counter-orbitally precessing tilted and warped accretion disk, the outer rim of 
which periodically blocks the line of sight to the X-ray emitting regions on the 
surface of the neutron star \citep[see, e.g.,][]{GerendBoynton76, 
HowarthWilson83, Shakura_etal99}. Roughly every $\sim 5$\,yrs Her~X-1 
exhibits so-called {\em anomalous low states} where the X-ray flux is 
strongly reduced and the 35\,d variability is seen only marginally 
\citep{Parmar_etal85, Vrtilek_etal94, Parmar_etal99, Staubert_etal09}.

Here we report on the timing analysis of X-ray data obtained  with
\textsl{RXTE}  (the \textsl{Rossi X-ray Timing Explorer}) and with
\textsl{INTEGRAL}. 
Combining these results with historical data from the literature 
we refine the global orbital ephemeris of Her~X-1 and rate of the secular 
decrease of its orbital period, which was originally reported by 
\citet{Deeter_etal91}. In addition, we were able to measure the eccentricity 
of the orbit for the first time (only upper limits had been reported up to date).


\begin{table*}
\vspace*{2mm}
\caption{Details of \textsl{RXTE} observations of Her~X-1 used for the timing analysis.} 
\label{obs_table}
\begin{tabular}{l l l l l l l l}
\hline\hline
35\,d cycle no. $^{1}$                       &  257           &  269              & 303            & 313             & 323            & 351                 &373      \\
  \hline\hline
Source of observational data            & RXTE         & RXTE           & RXTE         & RXTE          & RXTE         & RXTE/INTEGRAL & INTEGRAL \\
Date of observation                          & July 96       & Sep 97          & Dec 00       & Dec 01       & Nov 02       & July 05             & Sep 07 \\
Start of observation [MJD]                & 50290.027 &   50704.912  & 51896.319 & 52243.060  & 52600.061  & 53574.715      & 54346.088\\
No. of RXTE/INTEGRAL orbits used & 18              &  14                & 33              & 45               & 29               & 8 / 3                & 1       \\
Elapsed time of observation [days]   & 1.42           &  2.43             & 5.62           & 3.52            & 4.02            & 8.16                & 2.03   \\
Elapsed time of obs. [binary orbits]   & 0.83           &  1.43             & 3.30           & 2.07            & 2.36            & 4.80                & 1.20   \\
Coverage of binary orbits [\%]           & 36.6           &  10.0             & 10.0           & 30.7            & 18.9            & 37.0                & 53.0   \\
Number pulse profiles                       & 139            &  45                & 61              & 161             & 133             & 9 / 69              & 27      \\
Mean integration time per profile [s]  & 322            & 276              & 802             & 580             & 493             & 420 / 2350      & 3500  \\    
\hline
\end{tabular} 
\vspace*{1mm} \\
$^{1}$ The numbering of 35\,d cycles follows \citet{Staubert_etal83}:  
``Pulse profile counting" is applied, based on pulse profile shapes repeating with an average period of 34.98\,d \citep{Staubert_etal09}.
 \label{obs}
\end{table*}

\section{Observations\label{obs}}

Her~X-1 was repeatedly observed by the \textsl{Rossi X-ray Timing Explorer} 
(RXTE) \citep{Bradt_etal93} over a 10 years time frame (1996--2005). 
In addition, we make use of two extended observations by \textsl{INTEGRAL}
\citep{Winkler_etal03} in 2005 and 2007.
The observations provided high-quality X-ray spectral and timing data. 
For the timing analysis presented here we used data obtained with the
\textsl{RXTE Proportional Counter Array} (PCA) \citep{Jahoda96} in the energy 
range 3--20\,keV and with \textsl{INTEGRAL ISGRI} \citep{Ubertini_etal03} 
in the range 20--60\,keV. Seven sets of data, all taken during a \textsl{main on} state,
were long enough and sufficiently densely sampled such that a local ephemeris
could be determined (the set of 2005 is from partly simultaneous observations
by \textsl{RXTE} and \textsl{INTEGRAL}, see also \citealt{klochkov07, 
klochkov_etal08a,klochkov_etal08b}). We note, that the capabilities for relative 
and absolute timing required for the kind of analysis presented here are well met
by both \textsl{RXTE} \citep{Rots04} and \textsl{INTEGRAL} \citep{Walter03}.

Details of the observations are given in 
Table~\ref{obs}. Following the 35\,d cycle counting method adopted by 
\citet{Staubert_etal83}  and \citet{Staubert_etal09} the seven cycles are 
numbered 257, 269, 303, 313, 323, 351 and 373. Here ``pulse profile counting"
is applied, as defined by \citet{Staubert_etal09} (meaning that the changing shape
of the pulse profile is used to identify individual 35\,d cycles).
The PCA data were mostly taken in ``event  mode", providing arrival times 
for individual events, and a few times in ``binned mode", giving light curves 
(with time resolution, generally 1/16 s, sufficient to do timing analysis for the 
1.24\,s pulsation in Her~X-1).

\section{Timing analysis of \textsl{RXTE} and \textsl{INTEGRAL} data\label{timing}}

\subsection{Basic techniques to determine the pulse period\label{basic}}

We use two techniques for the determination of pulse periods: \textsl{epoch folding} 
with $\chi^{2}$ search \citep{Holt_etal76,Leahy_etal83} and 
\textsl{pulse phase connection}  \citep{Manchester77,Deeter_etal81}.
Both methods are generally well established and are appropriate in the case where 
the pulse period is already roughly known. Our data are either ``event lists" (the arrival 
times of single detector events) or ``light curves" (containing the numbers of 
events per second and the absolute times of the centers of those bins). 

With \textsl{epoch folding}, events are accumulated into bins of a phase histogram 
according to their phase with respect to an assumed trial period and a given
start time (e.g., the time of the first event), thereby generating a \textsl{pulse profile} 
(e.g., Fig. \ref{pp}, with our Her~X-1 data we generally used 128 bins). A search for 
the likely `best' period is conducted by constructing a $\chi^{2}$ distribution:
for any pulse profile generated with an assumed trial period  the value
$\chi^{2}$ = $\Sigma~(n_{i}~-\langle n \rangle)^{2}$/$\sigma_{i}^{2}$ is computed, where 
$n_{i}$ are the number of events in bin \textsl{i} and $\langle n \rangle$ is the mean of all
$n_{i}$.  $\chi^{2}$ is testing the deviation from a uniform distribution (which is
expected for a trial period far from the true one). The $\chi^{2}$ distribution 
is found by plotting the $\chi^{2}$ values for a series of trial periods (selected
to sample a sufficiently large period range with sufficient resolution). 
The `best' period is assumed to be the one leading to the pulse profile with the 
maximum deviation from a flat distribution. The $\chi^{2}$ distribution 
resembles a triangular distribution and can often be well fitted by a Gaussian,
the mean of which may be considered the `best' period (note that this is
likely better than using the maximum of the  $\chi^{2}$ distribution, especially
in cases of limited statistics). The full width at half maximum (\textsl{FWHM}) of 
the $\chi^{2}$ distribution should be of the order 
\textsl{FWHM} $\sim P^{2}/TOB$, if \textsl{P} is the true pulse period and 
\textsl{TOB} is the total elapsed time of the observation.

For \textsl{pulse phase connection} pulse profiles are produced, using the best 
estimate for the pulse period (e.g., from a $\chi^{2}$ search), for integration intervals 
of appropriate length (leading to profiles with sufficient statistics). The time differences 
between profiles of adjacent integration intervals (measured either by making use of
sharp features in the profile or by \textsl{template fitting}) are then identified with an
integer number of pulse cycles. In this way, a list of unique cycle numbers and 
associated absolute times are generated, which can then be modeled. 
Necessary conditions for the applicability of this method are: 
(1) the shape of the pulse profile must be stable enough for the common
details of different pulse profiles to be identified and (2) the time separations of 
the profiles must be small enough such that, with the given uncertainty of the
assumed pulse period, any mis-counting can be ruled out. If these conditions
are satisfied the \textsl{pulse phase connection} is by far the preferred
technique, allowing very flexible modeling, especially if the pulse period is
time variable, and with easy and accurate estimates of uncertainties. 

For a constant pulse (spin) period $P_{\rm s}$ the expected \textsl{arrival time} 
of pulse number \textsl{n} is
\begin{equation}
t_{\rm n} =t_{\rm 0} + n~P_{\rm s},
\label{linear}
\end{equation}
where $t_{\rm 0}$ is a reference time at the start of the observation.
For the case of non-zero first and second derivatives of the pulse period,
the \textsl{arrival times} are given by (see, e.g., \citealt{Kelley80,Nagase89})
\begin{equation}
t_{\rm n} = t_{\rm 0} + n~P_{\rm s} + \frac{1}{2}~n^{2}~P_{\rm s}\dot P_{\rm s} 
+  \frac{1}{6}~n^{3}~P_{\rm s}^{2}\ddot P_{\rm s} + \dots
\label{cubic}
\end{equation}
Here $P_{\rm s}$ is the initial pulse period valid at $t_{\rm 0}$.
Modifications for pulse arrival time delays, e.g. because of binary motion
of the pulsar, can easily be introduced (see below). Often Equ. (\ref{cubic})
is expressed in the inverted form giving the pulse phase $\phi$ at time 
$(t - t_{0})$ \citep{Blandford76,Manchester77,Kramer08}
\begin{equation}
\phi = \phi_{\rm 0} + \nu_{\rm 0}~(t - t_{\rm 0}) + \frac{1}{2}\dot \nu~(t - t_{\rm 0})^2
+  \frac{1}{6}~\ddot \nu~(t - t_{\rm 0})^3 + \dots
\label{linear}
\end{equation}
with $\nu_{\rm 0} = 1/P_{\rm s}$ 
being the spin frequency 
at the reference time $t_{\rm 0}$.

\subsection{Analysis with previous ephemeris\label{oldephem}}

Our initial approach to the timing data of Her~X-1 obtained with \textsl{RXTE}
was to determine accurate local pulse periods and to construct the corresponding
pulse profiles. The scientific motivation was two-fold: (1) to complete our data base 
on the evolution of the \textsl{pulse period} with time which shows a long-term 
mean spin-up of about 10\,ns per day. Superimposed to this trend, modulations 
are seen of relative spin-up and spin-down on different time scales, some of which 
appear to be quasi-periodic and correlated with the turn-on history of the 35\,d 
modulation due to the precessing accretion disk, e.g., \citet{Staubert_etal06a, 
Staubert_etal09}. (2) With the spin periods found in this way, pulse profiles were 
constructed in order to study the systematic variations in  \textsl{pulse shape} with 
phase of the 35\,d modulation. This study has in fact led to support the idea of two 
different, but linked 35\,d clocks in Her~X-1: the precession of the accretion disk 
and free precession of the neutron star \citep{Staubert_etal09}.
Further investigations with regard to the above two points are in progress.

\begin{figure}
  \resizebox{\hsize}{!}{\includegraphics{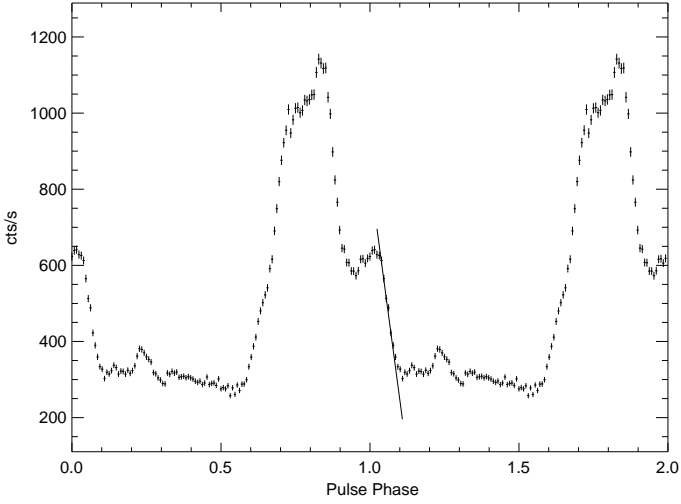}}
  \caption{An example of \textsl{PCA RXTE} pulse profiles used to study 
        pulse shifts due to the orbital motion of Her~X-1. The profile is from
        700\,s integration time at MJD 52243.786 (35\,d cycle no. 313). The solid line 
        shows the sharp decay after the soft trailing shoulder which was used
        as a phase reference (see text).}
  \label{pp}
\end{figure}

As a first step, all arrival times are translated to the inertial frame of the 
solar system barycenter. Then, the arrival times are corrected for
the binary motion of the X-ray emitting neutron star around its optical
companion, using the orbital elements and ephemeris of Her~X-1 given 
by \citet{Deeter_etal91}: $T_{\pi/2}$~=~MJED~43804.519980(14), 
$P_{\rm orb}$ = 1.700167720(10)~\,d, 
$\dot P_{\rm orb}$~=~$(-2.25\pm 0.27)\times 10^{-8}$\,$\rm d\,\rm yr^{-1}$,
$a\sin i$~=~13.1831(3)\,s, and eccentricity $\epsilon = 0.0$.
Next, integration intervals are selected, generally identical to the intervals 
of un-interupted observations (for \textsl{RXTE} the satellite orbits, for
\textsl{INTEGRAL} the so called Science Windows), for which the local
pulse period is determined by epoch folding/$\chi^{2}$ search and
the pulse profile is constructed. A typical profile is shown in Figure~\ref{pp}.
Next, the phase connection technique is applied to all profiles of the
groups of observations (Main-On states of Her~X-1).
Table~\ref{obs} gives details for the 7 Main-On states covered. 
This means that within each group every profile is associated with a
unique pulse number and an absolute arrival time (in MJD). To find the
latter we used a particular feature in the pulse profile of Her~X-1 as a time 
marker, namely the ``sharp edge" at the trailing edge of the right hand
shoulder of the main pulse going into a flat bottom minimum. This 
flat bottom is generally well defined, extending over $\sim 0.1$ in pulse 
phase over which the flux can be represented by a constant
(for the example profile shown in Fig.~1 it is from phase $\sim 1.1$ to 
$\sim 1.2$).  A straight line is fitted to the ``sharp edge" 
and the phase of the intersection between the straight line and the constant 
representing the minimum is determined. Since the ``sharp edge"
is only a few degrees from the vertical the systematic uncertainty introduced
by this procedure is quite small. The overall accuracy has been found to
be better than one bin for 128 bins per pulse period, that is $<10$\,ms. 
Combining this with the known absolute time for phase 0.0, i.e. the reference 
time at which the folding of the profile was started, generally the time of the first 
event, gives the arrival time assigned to this profile. We have selected the 
``sharp edge" as the time marker, because it appears to be the most stable 
feature of the pulse profile of Her~X-1, while most other features (including 
the main peak) change their shapes as a function of 35\,d phase. 
We note, however, that the timing results do not depend in any critical 
way on using  the ``sharp edge", consistent results have been obtained on the 
same data with template fitting using the complete pulse profile 
\citep{klochkov07,klochkov_etal08b}. The data sets used are short enough,
such that the shape of the pulse profile does not change significantly.
It is, of course, necessary that the accuracy of the assumed
pulse period is sufficiently high, such that no counting errors occur when 
bridging the existing gaps between data sets. 

\begin{figure}
  \resizebox{\hsize}{!}{\includegraphics{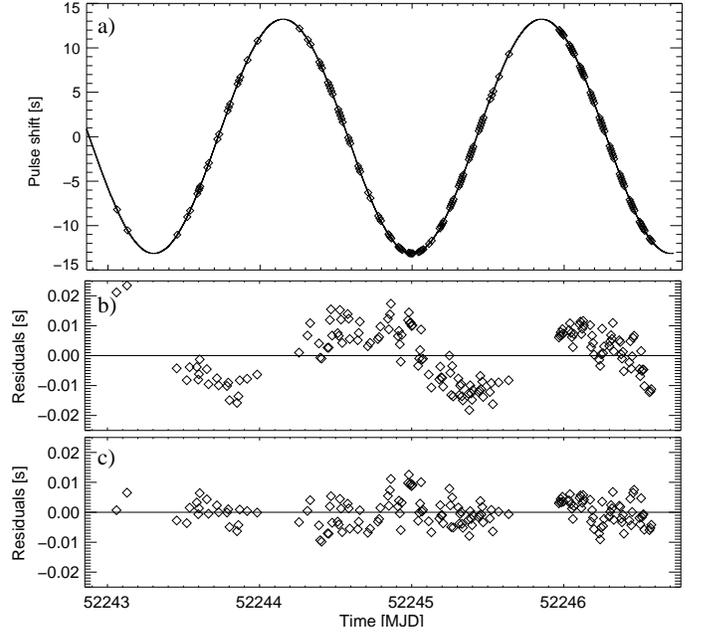}}
  \caption{a) Delays of the pulse arrival time in Her~X-1 due to its orbital motion
  and theoretical sine curve for 35\,d cycle no. 313 (Dec 2001). 
 b) Residuals to a fit using the orbital parameters based on \citet{Deeter_etal91}. 
  c) Residuals using the best-fit parameters for our analysis of non-binary corrected data.}
  \label{ephem}
\end{figure}

We like to mention here that the difference in energy range in which
the data have been taken by \textsl{RXTE} (3--20\,keV)   and by 
\textsl{INTEGRAL} (20--60\,keV) does not pose a problem for the current
timing analysis, also not for the overlapping observations of \textsl{RXTE} and 
\textsl{INTEGRAL} of 2005 July. First, each data set was analyzed 
independently, leading to independent, but consistent solutions for the local 
$T_{\pi/2}$ values. Since it is known that the pulse profiles of Her~X-1
(and other pulsars) change with energy, one might expect that the 
absolute time of the ``sharp edges" is energy dependent. Instead,
we have found that the 2005 data sets of both satellites could be combined 
into a common set. Any shift between the two sub-sets is less than five 
milli-second (well within the general uncertainty), supporting the notion
that the  ``sharp edge" is indeed a stable feature.

Applying Equ.~(\ref{cubic}), 
the observed pulse number/arrival time data were fitted and the best fit 
values for $t_{\rm 0}$, $P_{\rm s}$ and $\dot P_{\rm s}$ determined (with 
$\ddot P_{\rm s}\equiv 0$). A first set of $T_{\pi/2}$ values was obtained
which were combined with historical $T_{\pi/2}$ values from the 
literature to establish a revised global solution from which updated values for 
$P_{\rm orb}$ and $\dot P_{\rm orb}$ for any given time could be calculated.
However, we found that systematic sine-like residuals remained with an 
apparent period close to the orbital period, suggesting that the ephemeris
which we had used to correct the arrival times for binary motion was not
optimal. As an example, Fig.~\ref{ephem}b) shows the
observed residuals for 35\,d cycle no. 313 (Dec 2001). We therefore 
decided to perform a timing analysis (starting with  barycenter-corrected, 
but not binary-corrected arrival times) and to determine the local ephemeris
for all of our data sets. 

\subsection{Analysis of non-binary corrected data\label{non-binary}}

In the same way as described in Section~\ref{oldephem} the barycenter 
(not binary) corrected arrival times are used to generate sets of
pulse profiles and the corresponding lists containing pairs of pulse 
numbers and absolute arrival times. These lists were then taken for
fits with equ.~(\ref{cubic}) which was extended by the additional term
\begin{equation}
+~a\sin i\cos[2\pi~(t-T_{\pi/2})/P_{\rm orb}]
\label{cos}
\end{equation}
in order to describe the arrival time delays due to the changing position of 
the X-ray source along the binary orbit (assumed to be circular). $a\sin i$ 
is the projected radius of orbit in light-seconds and $T_{\pi/2}$ 
(=~$T_{\rm 90}$) in MJD is the 
time at which the mean orbital longitude of the neutron star is $90^\circ$,
corresponding to the maximum delay in pulse arrival time, or zero
pulse period derivative (in Her~X-1 essentially identical to the center
of eclipse). For these fits $a\sin i$~=~13.1831\,s \citep{Deeter_etal81,
Deeter_etal91}
and orbital periods $P_{\rm orb}$ (determined for each data set from the 
global solution found earlier) were kept fixed. The fit parameters,
determined for all data sets, are $T_{\pi/2}$, $t_{\rm 0}$, $P_{\rm s}$, 
$\dot P_{\rm s}$ ($\ddot P_{\rm s}\equiv 0$). 
For the data set of 35\,d cycle no. 313 (Dec 2001), the improvement is
evident in Fig.~\ref{ephem}c).

\begin{table*}
\caption{Orbital parameters of Her~X-1 determined from the \textsl{RXTE} 
observations during 35\,d cycles No.~257, 269, 303, 313, 323, 351 and
from \textsl{INTEGRAL} observations during cycles 351 and 373.$^{1}$
The uncertainties in parentheses are 1-sigma (68\%) ($\chi^2_{\rm min} + 1.0$) 
and refer to the last digit(s). The following
orbital elements were kept constant: a~sin~i = 13.1831\,s, omega = 96.0\,deg
and $\epsilon =  (4.2\pm0.8) \times 10^{-4}$. The values for $P_{\rm orb}$
are calculated using the best-fit orbital elements given in Table \ref{globsolut}.}
\label{ephemtab}
\centering
\begin{tabular}{l l l l l l }
\hline\hline
35\,d cycle no.$^{1}$  & $P_{\rm s}$~[s]  & $T_{\pi/2}$~[MJD (TDB)]  & $P_{\rm orb}$~[d]   & $dP_{\rm s}/dt~[ss^{-1}]$     & $dP_{\rm s}/dt/dt~[s^{-1}]$   \\
  \hline\hline
257                   & 1.237739612(25)         & 50290.659199(20)          & 1.700167399         & $-(2.8\pm1.0)~10^{-12}$       & 0.0  \\
269                   & 1.237729884(35)         & 50703.799869(50)          & 1.700167379          & $-(6.8\pm2.5)~10^{-12}$      & 0.0  \\
303                   & 1.237769831(12)         & 51895.617167(25)          & 1.700167321          & $+(1.54\pm0.17)~10^{-12}$  & 0.0  \\
313                   & 1.237767113(6)           & 52242.451312(10)          & 1.700167304          & $-(5.7\pm1.5)~10^{-13}$       & 0.0  \\
323                   & 1.237761690(4)           & 52599.486440(12)          & 1.700167287          & $+(2.8\pm0.2)~10^{-12}$      & $-(2.9\pm0.08)~10^{-17}$  \\
351                   & 1.237758409(8)           & 53573.682256(35)          & 1.700167240          & $-(7.0\pm0.8)~10^{-13}$       & 0.0  \\  
373                   & 1.237744750(60)         & 54345.558195(80)          & 1.700167202          & 0.0                                         & 0.0  \\ 
\hline
\end{tabular}
\vspace*{1mm} \\
$^{1}$ The numbering of 35\,d cycles follows \citet{Staubert_etal83}:  
``Pulse profile counting" is applied, based on pulse profile shapes repeating with an average period of 34.98\,d \citep{Staubert_etal09}.
 \label{results}
\end{table*}

\begin{figure}
  \resizebox{\hsize}{!}{\includegraphics[angle=90]{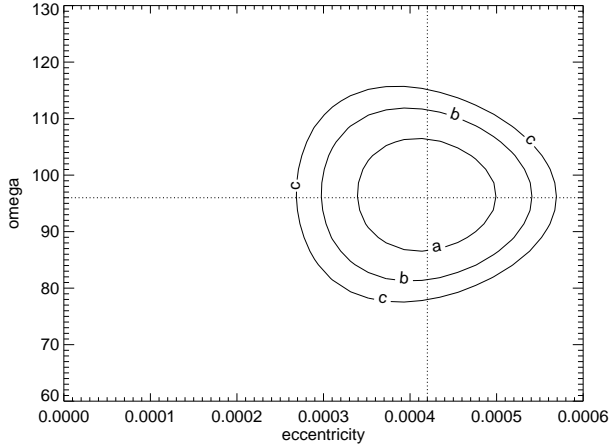}}
  \caption{Uncertainty contours for eccentricity versus omega
 from fits to the observations of 35\,d cycle no. 313. The error contours are for
 a: $\chi^2_{\rm min}+1.0$, b: $\chi^2_{\rm min}+2.3$, and c: $\chi^2_{\rm min}+3.5$. }
  \label{ecc}
\end{figure}

\subsection{Final fits solving Kepler's equation\label{Kepler}}

The results obtained so far allow us to establish a global solution with quite 
accurate values for $P_{\rm orb}$ and $\dot P_{\rm orb}$ for any time,
including the times for all of our observations.
Such a procedure is necessary since $P_{\rm orb}$ cannot be constrained 
from the individual data sets themselves. 

Keeping $P_{\rm orb}$ fixed at values calculated from the above global
solution, we have attempted to find a full orbital solution for each of
our individual data sets by solving Kepler's equation 
\citep{Blandford76,Nagase89}, in order to see whether 
we can constrain other parameters
which were so far kept constant: the projected orbital radius $a\sin i$, 
the eccentricity $\epsilon$, the longitude of periastron passage $\omega$
(in case $\epsilon$ is not zero) and the second derivative of the pulse 
period $\ddot P_{\rm s}$.

The first parameter tested was $a\sin i$. 
Our best data set (cycle no. 313, Dec 2001) gives
$a\sin i$~=~13.1831(5)~\,s. For all other data sets the values
are consistent with this (within uncertainties). A weighted average
of all values leads to $a\sin i$~=~13.1831(4)~\,s. The same value 
was found by  \citet{Deeter_etal81, Deeter_etal91}: $a\sin i$~=~13.1831(3)~\,s 
(also a weighted average of several sets of observations).
So we have used this value of $a\sin i$ as a constant input
parameter. We note here, that we had wondered about our inability
to reduce the uncertainty of this value, despite the fact that the
number of photons we have collected from Her~X-1 is about 20 times
larger than that collected by \citet{Deeter_etal81}. We suspect that
this may be due to the intrinsic variability of the pulse profile which
introduces additional scattering in pulse arrival times,
limiting the achievable accuracy independent of photon statistics.

The second parameter tested was the eccentricity $\epsilon$.
To our surprise, the fits of three data sets (35\,d cycle nos. 257,
313 and 323) yielded formal values for $\epsilon$ larger than zero.
For our best data set (cycle no. 313) a statistically significant 
value is found: $\epsilon$~=~$(4.2\pm 0.8)\times 10^{-4}$, the
corresponding value for the longitude of periastron is 
$\omega$~=~$(96\pm 10$) deg ($\chi^2_{\rm min} + 1.0$,
 innermost contour of Fig.~\ref{ecc}). This result suggests
 a significance of little more than 5 sigma. In order to further
 investigate the statistical situation, we have applied an F-test
 \citep{Bevington_69}: the reduction in $\chi^{2}$ 
 when adding $\epsilon$ and  $\omega$ as additional free 
 parameters is 27.2 at the final level of $\chi_{2}^{2}$~=~156.6
 with 156 degrees of freedom (dof). This leads to 
 $F_{\chi} = (\Delta \chi^{2}/\chi^{2}) / (\Delta(\mathrm{dof})/\mathrm{dof})$ = 13.5,
 corresponding to a probability of $<10^{-4}$ that the improvement
 in $\chi^{2}$ is simply due to statistical fluctuations. Further
 confidence in a significantly measured eccentricity is gained by 
 combining data sets.
We have combined the data sets of cycle nos. 313 and 
323 into one common set by correcting the arrival times for the 
local $\dot P_{\rm s}$ values and shifting the arrival times of cycle
323 to the absolute time frame of cycle 313, using our best
information on $P_{\rm orb}$ and $\dot P_{\rm orb}$. The 
fit to the combined data set yields 
$\epsilon$~=~$(3.8\pm 0.8)\times 10^{-4}$ and 
$\omega$~=~($82\pm 10$)~\,deg, consistent (within uncertainties)
with the results from cycle 313 alone. Adding the data from cycle 257
or cycle 304 leads again to consistent results, but no further improvement. 
Even though the combination of several data sets confirms and 
strengthens the results of the single data sets, we prefer to quote 
the results from cycle 313 alone as our measurement, since 
it results from the simplest and most straight forward analysis of our 
best data set. From here onward we have kept $\epsilon$ and
$\omega$ fixed at the values measured for cycle 313. 

Finally, we attempted to measure values for the second 
derivative $\ddot P_{\rm s}$ of the pulse period $P_{\rm s}$.
Only for one data set (cycle 323) a statistically significant value is
found, for all others  $\ddot P_{\rm s}$ is consistent with zero,
for cycle 373 also $\dot P_{\rm s}$ is consistent with zero
(see Table~\ref{results}).

All data sets have then finally been fitted, solving Kepler's
equation and leaving the following parameters to be determined:
$P_{\rm s}$, $T_{\pi/2}$, $t_{\rm 0}$, $\dot P_{\rm s}$
and $\ddot P_{\rm s}$. The final fit parameters and their
uncertainties are summarized in Table~\ref{results}. The 
uncertainties quoted are 1 $\sigma$ (68\%), determined from the 
projections of the ($\chi^{2}_{\rm min} + 1.0$) error contours onto
the parameter axis of interest, allowing all other interesting parameters
to vary during the error contour evaluation \citep{Lampton_etal76}.

The values of $T_{\pi/2}$ are significantly different from those 
calculated using the global solution of \citet{Deeter_etal91}. This is not 
surprising, giving the extrapolation over $\sim 20$\,yrs and the updated 
value for $\dot P_{\rm orb}$. In the next Section \ref{orbephem}
we will determine the updated global solution for $P_{\rm orb}$ and 
$\dot P_{\rm orb}$, using all available historical observations 
of the source combined with our new results provided in 
Table~\ref{results}.

\section{Orbital ephemeris\label{orbephem}}

\begin{figure}
  \resizebox{\hsize}{!}{\includegraphics{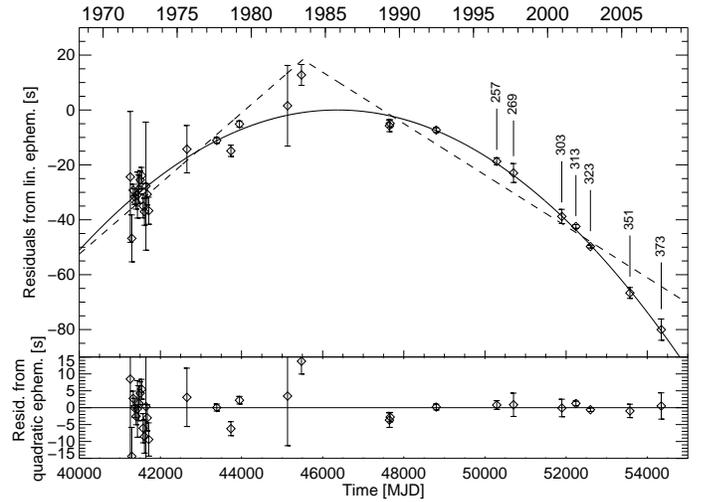}}
  \caption{Upper panel: Residuals of $T_{\pi/2}$ after subtracting the linear 
      part of the quadratic best-fit ephemeris. For the \textsl{RXTE} and
      \textsl{INTEGRAL} observations analyzed in this work 35\,d cycle 
      numbers are indicated. The solid line represents the best fit
      quadratic ephemeris. The dashed lines correspond to fits by the 
      linear ephemeris with a sudden jump of the orbital period around 
      1983. The lower panel shows the residuals to the best fit quadratic
      solution.}
  \label{ephemglob}
\end{figure}

Using our new values of $T_{\pi/2}$ (Table~\ref{ephemtab}) and
the historical data from previous missions 
\citep[][and references therein]{Deeter_etal81,Deeter_etal91,Wilson_etal94} we 
determine an improved value for the time derivative of the orbital period
$\dot P_{\rm orb}$ and an updated orbital ephemeris of Her~X-1 
valid for the whole $\sim$36\,yrs of observations since the discovery
of the source. Modeling all $T_{\pi/2}$ values (historical observations 
plus \textsl{RXTE} and \textsl{INTEGRAL} data) by a linear ephemeris 
($\dot P_{\rm orb}\equiv 0$) results in a very poor fit: 
$\chi^2_{\rm red}~=~58.9$ for 28 d.o.f. 
Thus, the linear ephemeris can clearly be rejected.
Figure~\ref{ephemglob} (upper panel) shows the residuals of $T_{\pi/2}$ 
after subtracting the linear part of the quadratic best-fit ephemeris.
For the quadratic ephemeris ($\dot P_{\rm orb}\ne 0$, shown by the solid curve) 
the fit gives $\chi^2_{\rm red}  = 1.8$ for 27 d.o.f.. If the outlying data point 
around MJD $\sim$45000 (due to {\em Tenma}, \citealt{Deeter_etal91}) which 
deviates from the quadratic fit by more than 3$\sigma$ is removed, 
$\chi^2_{\rm red}$ is reduced to 1.3.

The best fit parameters for the 
quadratic ephemeris are listed in Table~\ref{globsolut}. $P_{\rm orb}(0)$ 
is the orbital period at the reference time $T_{\pi/2}(0)$. For any 
particular observation $T_{\pi/2}$ and $P_{\rm orb}$ can be found using
the following formulae:
\begin{equation}
P_{\rm orb}(t) = P_{\rm orb}(0) + \dot P_{\rm orb}(0)\cdot(t-T_{\pi/2}(0)),
\end{equation}
\begin{equation}
T_{\pi/2}(n) = T_{\pi/2}(0) + n~P_{\rm orb}(0) + \frac{1}{2}n^{2}P_{\rm orb}(0)\cdot\dot P_{\rm orb}(0);
\end{equation}
where $n$ is the orbital cycle number ($n=0$ corresponds to 
$T_{\pi/2}(0)$). 

\citet{Deeter_etal91} had also discussed an alternative model, representing
constant orbital periods before and after a sudden change of the orbital period 
in 1983 (at MJD $\sim$45000). This appears to be mainly  driven by the 
outlying {\em Tenma} data point. We have performed linear fits to
the data before and after 1983, they are shown as 
dashed lines in Figure~\ref{ephemglob}). While the data set before 1983
is marginally consistent with a linear function ($\chi_{\rm red} = 1.8$ for 
18 d.o.f.), the data set after 1983 is clearly not: $\chi_{\rm red} = 9.1$ for
8 d.o.f.). We, therefore, reject the idea of a sudden change of the orbital 
period in 1983.

\section{Discussion\label{discuss}}

The value of $\dot P_{\rm orb}=-4.85(13)\times 10^{-11}$~$\rm ss^{-1}$
determined in Sect.~\ref{orbephem} and its uncertainty are significantly 
smaller than the corresponding values determined by \citet{Deeter_etal91} 
($-6.16(74)\times 10^{-11}$). In order to connect the derived value of the 
secular variation of the orbital period with the physics of mass exchange in the 
system, \citet{Deeter_etal91} had considered a very simplified model, in which 
the two stars comprising the binary are treated as point masses and only 
orbital angular momentum is considered. In the case of conservative
mass transfer one can obtain:
\begin{equation}
\frac{\dot M_{\rm ns}}{M_{\rm ns}}-\frac{\dot M_{\rm ns}}{M_{\rm opt}}+
\frac{\dot P_{\rm orb}}{3P_{\rm orb}}=0,
\label{simpmod}
\end{equation}
where $M_{\rm ns}$ and $M_{\rm opt}$ are the masses of the neutron
star and the optical companion, respectively, and $\dot M_{\rm ns}$ is the
accretion rate of the neutron star. In this idealized case, using 
$M_{\rm ns}=1.4M_\odot$, $M_{\rm opt}=2.2M_\odot$, and the
value of $\dot P_{\rm orb}=-4.85\times 10^{-11}$~s/s, we obtain a mass 
accretion rate of the neutron star of 
$\dot M_{\rm ns}\sim 1.3~10^{-8}\,M_\odot~\rm yr^{-1} \sim 8\times 10^{17}\,{\rm g~s}^{-1}$.
This value is slightly smaller than that found by \citet{Deeter_etal81} but 
still roughly an order of magnitude higher than the accretion rate necessary
to provide the observed X-ray luminosity $L_{\rm X}\sim 2\times 10^{37}\,
{\rm erg~s}^{-1}$ (assuming that the radiative efficiency of accretion
is $\sim$10\%). To explain this \citet{Deeter_etal81} had suggested that the 
mass transfer is not conservative ($\dot M_{\rm ns}\neq -\dot M_{\rm opt}$) 
and matter may leave the system in the form of magnetically channeled 
wind from the X-ray heated surface of the companion star 
\citep[see][]{Ruderman_etal89} carrying away angular momentum. 
We argue, however, that the model used for conservative mass transfer 
is very approximate. It does not take into account  the angular momentum 
of the accretion disk and the accretion stream, their dynamic action on the 
stars as well as the shift of the center of mass of the system during the mass 
transfer, which are known to be significant factors influencing the final result 
\citep[see e.g.][]{DSouza_etal06}. Furthermore, one might expect exchange
of angular momentum by spin-orbit coupling \citep{Marsh_etal04}. Thus, 
Equ.~(\ref{simpmod}) is very approximate and may provide only a rough 
estimate (to the order of magnitude) of $\dot M_{\rm ns}$. The discrepancy 
we observe, therefore, cannot rule out a conservative mass transfer scenario.

\begin{table}
\caption{Orbital elements of Her~X-1. The uncertainties in parentheses (68\%) 
refer to the last digit(s). Note that \citet{Deeter_etal81} had measured
a~sin~i  = 13.1831(3)\,lt-s (with a slightly smaller uncertainty). All other values
are updated or new.
}
\label{globsolut}
\centering
\begin{tabular}{l l l}
\hline\hline
$T_{\pi/2}(0)$~[MJD (TDB)]  &=& 46359.871940(6) \\
$P_{\rm orb}(0)$ ~[d]                    &=& 1.700167590(2)  \\
$\dot P_{\rm orb}(0)$~[d/d]           &=& $(-4.85\pm0.13)\times 10^{-11}$    \\
$a~sin~i$~[lt-s]                              &=& 13.1831(4) \\
eccentricity $\epsilon$                   &=& $(4.2\pm0.8) \times 10^{-4}$ \\
omega~[deg]                                 &=& $96.0 \pm 10.0$   \\
\hline
\end{tabular}
 \label{values}
\end{table}

The alternative to a continuous decrease of the orbital period, namely a 
sudden change of the orbital period around 1983 as discussed by 
\citet{Deeter_etal81}, had received observational support by the analysis 
of \citet{Stelzer_etal97}, who performed a fit of the timing data of
Her~X-1 available at that time, finding that the model with a sudden jump 
provided a smaller $\chi^2$ than the model assuming continuous decrease 
of $P_{\rm orb}$. Our current observational result clearly rejects this 
conclusion (Sect.~\ref{orbephem}). This is reassuring in the light of several 
theoretical difficulties which had been discussed by \citet{Deeter_etal81}.
Here we add that from the disk--stream coupling model described in
\citet{Staubert_etal09} we expect a decrease of the mass transfer to the 
neutron star during \textsl{Anomalous Lows} (AL), not an increase as
assumed by \citet{Deeter_etal81}. This is supported by the observed 
spin-down during most of ALSs \citep{Staubert_etal06a}. 
Thus, we argue that the decrease of the orbital period of 
\hbox{Her~X-1/HZ~Her} originally discovered by \citet{Deeter_etal81} 
is due to a constant negative $\dot P_{\rm orb}$ and is not confined 
to any short time interval (like a jump around 1983) and that the most 
probable explanation is a continuous mass transfer from HZ~Her 
to the neutron star in a conservative or non-conservative scenario.

\section{Summary and conclusions\label{concl}}

The precise timing analysis of the {\sl RXTE} an \textsl{INTEGRAL} 
observations in combination with the historical data allowed us to improve 
the value for the secular change of the orbital period in Her~X-1/HZ~Her, 
originally reported by \citet{Deeter_etal81}. Our absolute value of 
$\dot P_{\rm orb}$ is somewhat smaller than that determined by 
\citet{Deeter_etal81}, and the uncertainty is significantly less. We
conclude that we cannot rule out the conservative mass transfer scenario. 
The data rule out the possibility of a sudden jump of the orbital 
period around 1983 as an alternative for a continuous decrease. The new 
value for $\dot P_{\rm orb}$ along with the values of $T_{\pi/2}$ and 
$P_{\rm orb}$ (listed in Table~\ref{globsolut}) establish a new
orbital ephemeris of Her~X-1 that can be used to calculate the orbital 
period, $P_{\rm orb}$, and the time $T_{\pi/2}$ for any particular time. 

A further result is the first measurement of a non-zero eccentricity
$\epsilon = (4.2\pm 0.8)\times 10^{-4}$ of the Her~X-1/HZ~Her orbit.
For most practical purposes, such a small deviation from zero may not
be relevant. However, in searches for subtle effects, e.g. a precession
of the plane of the orbit, possibly associated with precession of HZ~Her
\citep{Deeter_etal76b,Deeter_etal91}, such accurate values may be needed.

\begin{acknowledgements}
We acknowledge the support through DFG grants Sta 173/31 and
436 RUS 113/717/0-1 and the corresponding RBFR grants 
RFFI-NNIO-03-02-04003 and RFFI 06-02-16025, as well as 
DLR grant 50 0R 0302. We thank L. Rodina for her endless
efforts with respect to the data reduction, K. Postnov for
discussions about (non)conservative mass transfer in binaries
and R. Rothschild for valuable comments to the manuscript.
We thank the anonymous referee for relevant questions and valuable
suggestions.
\end{acknowledgements}

\bibliographystyle{aa}
\bibliography{refs}

\end{document}